# A Synthesizer Based on Frequency-Phase Analysis and Square Waves[1]

**Sossio Vergara**
Facultad de Ingeniería
Universidad ORT, Montevideo, Uruguay
ITI B. Pascal, Rome, Italy

## *Abstract*

The generation of synthetic signals has been one of the first applications of computers. As a matter of fact the earliest electronic computers were analog and their output was, in effect, a signal. With the advent of digital electronic the initial method employed has been the application of the Fourier Theorem, generating signals as series of sinusoids, a technique deserving a name by its own: "additive synthesis". The advantage of the technique is the great control on the parameters of the generated wave. The main disadvantage is the complexity of the computation involved, namely for each component many samples of sinusoid need to be computed, and this usually requires a special hardware to be performed in real time. The reason is that the sine wave, although being natural for physical linear systems, is very complex in the digital domain. This article introduces an effective generalization of the polar flavor of the Fourier Theorem based on a new method of analysis. Under the premises of the new theory an ample class of functions become viable as bases, with the further advantage of using the same basis for analysis and reconstruction. In fact other tools, like the wavelets, admit specially built nonorthogonal bases but require different bases for analysis and reconstruction (biorthogonal and dual bases) and vectorial coordinates; this renders those systems unintuitive and computing intensive. As an example of the advantages of the new generalization of the Fourier Theorem, this paper introduces a novel method for the synthesis that is based on frequency-phase series of square waves (the equivalent of the polar Fourier Theorem but for nonorthogonal bases). The resulting synthesizer is very efficient needing only few components, frugal in terms of computing needs, and viable for many applications.

---

## 1. Introduction

There is often the necessity to generate a specific analog signal, at frequencies ranging from audio to RF. This problem has been tackled in different ways during the evolution of technology, from analog means to digital computers. Different methods have been employed, from recording on magnetic tape to sophisticated algorithms [1]. All these methods share some common characteristics: they must be flexible enough to allow some degree of manipulation of the output and they must be compatible with the technology of the time. The first requirement is self evident, one usually needs to generate signals, generically based on a prototype wave, but wants some degree of freedom to adjust the output to tailor specific needs, maybe changing the pitch, filter or add components and so on. The other feature means that the technology is driven by the knowledge. A given method is used until advances in our knowledge makes its technology obsolete. Since the advent of digital computers many algorithms have been devised to use the new technology as a signal generator. One of the first of such algorithms is based on the Fourier Theory: as any signal can be decomposed into a series of sinusoids, so one can add sinusoids to generate any signal [1]. The main advantage of the method is that it is intuitive: it is easy to imagine beforehand the effect of a change on the harmonic content of a signal. The sinusoid appeared at the beginning a natural choice as it is the *eigenfunction* of all linear systems, so elemental and quite common in nature. Unfortunately the sinusoid is not easily generated by digital means, needing complex computation or at least a long Look Up Table (or wavetable) to store the values of the sine in function of the angle. Consequently specialized hardware must be installed to manage the calculations when real time synthesis is required for special applications, at fast sampling rate or whenever the available computing power is scarce. More on the architecture of a typical Fourier–based synthesizer below. Due to these difficulties other methods were developed that could be less computing intensive [1]. From subtractive synthesizers, essentially bank of filters (used mostly for speech), to Frequency Modulation, wavelets, and recently PCM (just storing digital samples of the signal, a technique made possible by the decreasing cost of memories) [1]. Any of these techniques has its own shortcomings, oscillating from being not flexible enough to being very computing intensive. The ideal still would be a technique that has all the flexibility of the Fourier algorithm but with low computing complexity and possibly not needing special hardware, so that it can be used in real time on general purpose processors, even in the case of signals at a frequency much higher than the audio band.
This is the goal of the paper: to introduce a new analysis-synthesis technique, flexible yet computationally not demanding, as to be carried out in real time even on a very simple architecture. Essentially what we are looking for here is a generalization of the Fourier Theorem in polar coordinates.
There have been previous works generalizing the Fourier Theorem: first to other orthogonal functions as the Legendre polynomials, and successively it has been extended to wavelets and frames [7] [9]. Lately it has been demonstrated that some special pairs of even and odd periodic functions, among which the square waves, can be used in frequency analysis [2] [12] [13]. As a matter of fact in [15] there is the example of a sinusoid generated as a sum of even and odd square waves. But the theory developed there, like the others in the field of nonorthogonal bases like wavelets and frames [7] [9], involve the use of biorthogonal and dual bases, and vectorial decomposition. The use of different bases for analysis and synthesis is a limiting factor for these tools. It is evident that when manipulating a signal it is more intuitive to understand the effect of adding or subtracting components (filtering), if the same basis is employed in both the analysis and the synthesis. Moreover the architectures based on these techniques are generally complex and computing intensive, and the mathematics involved limits the possibilities to vectorial decomposition only.

We think that the ideal tool for digital signal synthesis should employ square waves. The use of polar (or more generally: phase) decomposition, as opposed to the vectorial one found in all the previous works, should be enforced as to halve the computing demand and to make the method more intuitive.

The advantage of phase decomposition results from the fact that the basis is composed by a single function, instead for vectorial synthesis two components of the basis must be manipulated at each frequency.

Furthermore the analysis should be straightforward and use the same basis for analysis and synthesis so as to avoid dual and biorthogonal bases. It seems a bold statement but it is exactly what this paper is all about. In short, the trick is changing our approach to the analysis phase. The main impediment to the use of nonorthogonal bases is in fact the choice of the analysis method, namely the use of the inner product that is limited to orthogonal projections. Consequently for the computation of the components, biorthogonal and dual bases are necessary when nonorthogonal bases are involved (biorthogonal bases can be made orthogonal if needed).

On the other hand, a basis is defined as: "a set of functions that is capable of reconstructing any function of the given space". The analysis is not mentioned in the definition so, as long as we can reconstruct any function in terms of the given basis, we have complete freedom in the analysis phase. And we shall exploit this freedom using a novel iterative approach for the analysis, one that is centered on the specific reconstruction algorithm and the given basis.

The theory of frequency decomposition on an ample class of nonorthogonal bases has been already introduced in [11], but there it was limited to vectorial bases (like our standard Fourier tool, where a basis is a complex exponential or a sine cosine pair).

The theory in [11] is based on a completely new approach to the analysis as opposed to the established methods of wavelets, frames and generic frequency decomposition [7] [9] [2] [12] [13], overcoming the limitations of these traditional tools.

We stress again that the ideas developed here are not an evolution of the existing methods, instead this paper is just on a special application of a newer tool. The complete theory has many widespread consequences that will be discussed in a forthcoming paper.

In order to make the paper more readable, the mathematical details are confined in the appendix. Actually the appendix has a broader scope demonstrating that, once we use the new recursive computation, any periodic function $S(kx) \in L^2$ that satisfies quite loose requirements is a basis of the $L^2$ space (the space of Lebesgue square integrable functions). The square wave happens to be just one of the infinite possible bases. However in this paper, as a test application, we shall focus on the advantages that such approach can have on the sound (or broadly signal) synthesis, leaving the general discussion for another article. Now that we know how to decompose any signal, not just as a sum of sinusoids, but as a series of other functions, we can generate richer sounds adding only few instances of a complex basis instead of many simple sinusoids. Indeed, it is much more efficient to synthesize a complex sound starting from a basis that has some similarity with it. The consequence on the computing power needs is evident.

The procedure sports some similarity with wavetable synthesis [3]. There a period of the actual sound is stored in memory as a way to reduce the computing burden. Instead, using the new tool, we can employ in the wavetable any of a wide range of waves, and use the analysis developed here to compute the components of any sound as a series of the new basis (the recorded wave). For example one could imagine the possibility of synthesizing a musical instrument as a series of waves recorded from another, completely different, instrument. Just changing the number of components of the series one could continuously modify the sound from one instrument to the other.

But we chose here to explore the opposite direction: we shall look for a way to simplify the hardware of a real time additive synthesizer, so that much cheaper systems could be feasible, without sacrificing any of the features of the Fourier additive synthesizer. And the way to do it is to put the square wave at work. In fact, while it is not a problem to synthesize a sound in real time with current hardware, the actual methods are of limited use when very high sampling rate is required or when computing resources are scarce, for example when needing signal synthesis or processing on low power and/or low cost architectures. Please note that the sound synthesis is just one of the many possible applications of the theory developed here. Other applications are easily foreseeable exploiting the same mathematics, and few examples are given in the last section.

## *2. The square wave as a basis*

To be convinced that the square wave is a viable basis one needs simply to invert the Fourier expansion for that function. It is well known that if $w(\omega)$ is a square wave then:

$$w(\omega) = \sum_{n=0}^{\infty} \frac{1}{2n+1} \cos(\omega(2n+1)) = \cos(\omega) + \frac{1}{3}\cos(3\omega) + \frac{1}{5}\cos(5\omega) + \frac{1}{7}\cos(7\omega) + \frac{1}{9}\cos(9\omega)\ldots \quad (1)$$

From this equation we can subtract a square wave of amplitude one third and frequency three, this way all the sinusoids at a frequency multiple of three disappear:

$$w(\omega) - \frac{1}{3}w(3\omega) = \cos(\omega) + \frac{1}{5}\cos(5\omega) + \frac{1}{7}\cos(7\omega) + \frac{1}{11}\cos(11\omega) + \frac{1}{13}\cos(13\omega)\ldots \quad (2)$$

Now, subtracting a square wave of amplitude one fifth and frequency five from the (2), other components vanish but the component at frequency fifteen reappears with the minus sign:

$$w(\omega) - \frac{1}{3}w(3\omega) - \frac{1}{5}w(5\omega) = \cos(\omega) + \frac{1}{7}\cos(7\omega) + \frac{1}{11}\cos(11\omega) + \frac{1}{13}\cos(13\omega) - \frac{1}{15}\cos(15\omega)\ldots \quad (3)$$

So, repeating the procedure, one component at a time, one can isolate the fundamental sinusoid by means of a series of square waves. At each new square wave component the error gets reduced and moved to higher frequencies, just as happens with the usual Fourier reconstruction. And if one can reconstruct a sinusoid via square waves, thanks again to the Fourier Theorem, it means that one can reconstruct any other function $f \in L^2$ as a series of square waves. So the square wave is a basis for $L^2$. Naturally, a more direct algorithm exists for the computation of the square waves components of any function as indicated in the Appendix. And once we get here there is no reason why we should limit ourselves to the square wave. In the appendix the mathematical justification is generalized to an entire class of functions of which the square wave is just a special case. Here we shall focus only on the square wave basis, leaving the general discussion to a forthcoming paper.

In Fig. 1 a sinusoid has been reconstructed by means of 21 square waves, up to a frequency 55 times the fundamental. One can notice that the noise (in the form of dithering) in Fig. 1 is at a much higher frequency than the highest component (at frequency 55) and that the noise is higher at the peaks of the sinusoid.

To make things more clear, in Fig. 2 a sinusoid with its square wave reconstruction is plotted, but the approximation is now limited to the first three components at frequency one, three and five. The reconstruction reveals components at a frequency higher than five (most evident in the two "horns" at the peaks) and the noise is clearly higher there.

This effect is the result of the interaction of the components at frequency three and five whose periods are not integer multiples (see the bottom plots in Fig. 2), and it is due to the nonorthogonality of the bases. Their sum introduces the spikes at frequency fifteen at maxima and minima in Fig. 2. This is the origin of the very high frequency noise in Fig. 1 where one would expect instead a staircase type of reconstruction at frequency fifty-five. The highest frequency of the spikes in the reconstruction is generally determined by the product of all the prime frequencies of the components (that are nonorthogonal to each other), whereas of course in real synthesizers the spikes duration can be at the most equal to the sampling interval. This type of noise can be easily eliminated digitally, especially when employing oversampling. Even more surprising is that, when synthesizing special functions like the sawtooth, needing only even square wave components, this type of noise is absent.

It is interesting to notice that this is somehow the opposite of the Gibbs' phenomenon, suggesting that it must be seen as an aspect of a more general process. The Gibbs' phenomenon is a ringing at the discontinuities of a square wave built as Fourier series. It is due to the rather slow convergence of the Fourier series to the square wave, and it is related to the fact that the sine is a continuous function and is incapable of reproducing discontinuities (as already Lagrange pointed out) [4][5]. Exactly the opposite happens for the square wave basis that is a discontinuous function and has difficulties in reproducing a function when the derivative is lower. However the noise in Fig. 1 is at high frequency and can be eliminated with a simple low pass filtering, or even a crude digital averaging, leading to a very good reconstruction of the sinusoid. On the other hand, the Gibbs' effect on a square wave built by a Fourier series, can be eliminated only at the cost of degrading the discontinuity. In this respect the square wave series can be superior to the Fourier series.

Is its nature that renders the square wave more efficient in the reconstruction of signals with fast transients. As a matter of fact what changes faster than a square wave? And a square wave in this approach is synthesized by only one component, just the opposite of Fourier theory. The topic has many interesting implications but for lack of space we leave a more detailed discussion to a next forthcoming article. Here we just point out the fact that this new tool should be considered as the generalization of the Fourier theory, hence an entire new framework needs to be developed.

In case the sinusoid seemed too elemental to prove our claim, in Fig. 3 there is a more complex signal obtained as a sum of square waves. In Fig. 3 the signal has been approximated with just nine square waves. If we consider that the original signal is the sum of seven Fourier components (harmonics), and extends up to a frequency eleven times the fundamental, the approximation is surprisingly good. In Fig. 4 the same signal has been better approximated by summing thirty six square waves, up to a frequency that is fifty times the fundamental. Always in the figures, dotted curves represent the original signal, while solid lines the reconstruction. One can see that in Fig. 4 the noise is lower and shifted to higher frequency as it should be in any well behaved reconstruction.

Referring to Fig. 5, the "square wave frequency spectrum" is shown, which is the plot of the amplitude $M_k$ of the square waves as a function of frequency, relative to the approximation of Fig. 4.

A corresponding plot of the phases can be produced. At this point one could even think to modify this "generic spectrum" as the one in Fig. 5 to get some kind of "generic filtering". The result on the reconstructed signal would be at first absolutely counter-intuitive but nonetheless interesting. Briefly, as Fourier analysis models Linear Systems, so the analysis in nonorthogonal frequency domain (with nonorthogonal functions as basis) can be employed for Nonlinear Systems. Using nonorthogonal bases in frequency domain is possibly an efficient way of performing Nonlinear Signal Processing. Details will be given in a subsequent paper.

At this point it could be interesting to compare the reconstruction of a sinusoid obtained by means of this tool, and by means of wavelets, as for example the Haar set [6], that is generated starting right from a square wave, via the usual operations of shift and scale, as shown in Fig. 6.

As can be seen, the wavelet reconstruction of the sine in Fig. 7 is poorer, in the sense that the RMS noise is higher, despite in Fig. 1 we used less components (21 square waves as opposed to 32 wavelets). Furthermore the extra noise in Fig. 1 is at much higher frequency and easily eliminated as discussed above. To be less biased the situation is not as bad as it appears for the wavelets, in fact the square wave decomposition needs two parameters for each component: amplitude and phase, whereas for the wavelets only the amplitude is required. So the resulting data reduction is less favorable to the square waves as could appear from the comparison of the Fig. 1 and Fig. 7. Nevertheless even taking into account the double set of coefficients, the square wave reconstruction results more accurate and easier to deal with (only two values +1,-1 as opposed to three for the Haar wavelet).
The wavelets have indisputable advantages over the standard frequency decomposition in their capability to perform a multirate analysis and in being time localized [7][8][9]. However, the wavelet analysis – synthesis tool has some drawbacks: only few sets of wavelets are available, and they are quite poor as reconstruction medium, requiring a greater number of components, and higher computing power in order to be usable in real time [10] as compared to square waves.

## *3. Architecture of the square wave additive synthesizer*

To see what the architecture of a square wave based synthesizer would be, we first look at a typical additive synthesizer based on Fourier synthesis as the one sketched in Fig. 8. As it is sinusoid based, it needs a way to compute the sinusoids, the typical solution is to introduce a Look Up Table (LUT) memory and eventually a mechanism for the interpolation of samples not contained in the LUT. The synthesizer is essentially a series of counters of frequency "Step" and phase "Pha" incremented at each cycle and used to address the LUT, the output of which is the actual sample of the sinusoid. Once a sinusoid sample has been computed it must be multiplied by a given amplitude, that is essentially the modulus of the Fourier series expansion for the signal (stored in the Amplitude memory).
Finally all these values must be accumulated in the final adder to produce a single sample of the synthesized signal.
Now let us simplify the previous architecture to accommodate for the square wave basis. We do not need the LUT as the square waves are purely digital, and, as we associate the two levels of the square wave to the values of +1 and −1 we do not even need a multiplier, a simplest 2's complement logic will be sufficient. So the architecture for the square wave based synthesizer would be as in Fig. 9. The square waves are simply the most significant bit (msb) of the counters; it controls a 2's complement logic that is the equivalent to a multiplication by +1 or −1 of the amplitude. These values are finally accumulated to produce a sample of the desired signal.
We can see that only two adders and some memory space are needed in order to synthesize any waveform in real time. It means that even very cheap devices can generate complex signals. The estimated silicon area saving is around 70 - 80%. Looking at high end applications like music synthesis, one can see that, in order to reduce the high frequency noise associated with square wave synthesis as in Fig. 4, one could include few more square waves at higher frequency. Hence one can establish a sampling rate higher than that required for the Fourier synthesis, and so easing the job of the final low pass filter (essentially oversampling).
For instance, an ideal (Fourier) low pass filter at half the sampling frequency would eliminate all higher Fourier components, so reducing a square wave at the cut-off frequency to a plain sinusoid. As ideal filters, as all ideal things do not exist in this our Aristotelian world, all one has to do is to double or triple the sampling rate, and use a real low pass filter. It is equivalent to reconstructing the highest frequency sinusoid as a sum of two or more square waves.
In effect being the square wave the natural output of digital systems, its implementation is very cheap, so increasing the sampling rate is not a problem. One can reproduce acoustic sounds using high sample rate on very simple circuitry, employing many square waves. That is not the case when standard techniques are involved as additive synthesis or PCM. There a more complex architecture is needed that in turn limits the possibility of frequency increase.

This leads us to a further enhancement to the architecture of Fig. 9. There the computation is carried out at any cycle, i.e. to generate any sample of the sound all the square wave components must be added.

Let us suppose we are generating 2 channels (left and right) at 100KHz of sampling rate, then we have 5 microsecond of computing time for each channel. If we have 100 square wave oscillators for each channel, it means that we have 50 nanoseconds of computing time for each square wave oscillator. Almost any modern microprocessor can do the job.

But we can further improve the procedure.

The fact is that only the highest frequency square wave changes every cycle, the lower frequency square waves remain constant for many of the clock cycles. It is an additional advantage with respect to the sine wave additive synthesizer. From the bottom part of Fig. 2 one can see that for most of the time only few of the square waves change sign in the same cycle; while the rest remain constant. So it is useless to sum them every cycle. A sort of differential technique can be used: the output is kept constant and only when one of the square wave switches, the output is updated for the relative factor. This approach is best suited for software implementations, in which a processor can time share synthesis and other jobs. An even simpler version, when using only few square wave oscillators, can be designed using just an adder, few counters and some registers.

It can be seen that a synthesizer based on square waves is a viable alternative in many different applications, as a matter of fact it has the advantage with respect to the sinusoid additive synthesizer of allowing many possible implementations.

When very high frequency signals are needed, a completely custom integrated circuit is viable; in this case the output frequency can be foreseen in the order of hundreds of MHz, so allowing for digital real time synthesis in a field where the only viable implementations were, up to now, analog or Direct Digital Synthesis (DDS). DDS is used for high frequency signals and employs a look up table (similar to PCM technique used for music) storing an entire period of the signal. The LUT can be read at high speed, but can be updated at lower speed, making the system not truly real time. Instead, for signals in the audio band, a dedicated DSP processor or even a general purpose microprocessor can be employed, thus greatly reducing the cost of such instruments, while retaining the effectiveness of the additive synthesis.

This approach can thus effectively lead to software only, real-time synthesis of complex sounds on standard personal computers.

Finally, very simple, low end implementations are possible, so allowing this technique to be promptly used in a large class of devices. Experimental devices have been created by students at the ORT University in Montevideo, Uruguay using an FPGA. One has an architecture exactly like the one presented in Fig. 8; the other is even more simple: just a handful of programmable counters, few registers, a 2's complement and an adder constitute the synthesizer.

Both have shown satisfactory audio capabilities, needing only a DAC and a plain RC filter at the output in order to generate low noise signals.

Furthermore, the same technique can be easily extended to data transmission and compression and to multidimensional signals as for example images. The current methods for image compression involve the use of DFT, DCT or DWT (Discrete Transform of Fourier, Cosine or Wavelet types). These algorithms are symmetrical, the direct and inverse have the same computational complexity. But this is not the optimal case because data (i.e. an image) is compressed only once and decompressed every time is viewed. In these cases an asymmetrical algorithm is "ecologically" preferable. And this is exactly the case of frequency decomposition on square waves. The analysis (or generally compression) algorithm is twice the complexity of the FFT, but the synthesis (decompression) is algorithmically very simple. Moreover the most promising image compression algorithm is based on Haar wavelets and above has been shown that the polar frequency decomposition in square wave basis can be more efficient than the Haar wavelets, needing a lower number of components for the same noise level and a simpler algorithm.

## *Conclusions*

The goal of the article was to introduce a generalization of the Polar version of the Fourier Theorem having widespread consequences, and complementing a previous work [11] dedicated to nonorthogonal vectorial bases. To put to the test the possibilities of the new tool, an example application was chosen as test bench. It has been demonstrated that, when employing a novel recursive method for the analysis, a large class of functions, including the square wave, become viable bases for the $L^2$ space. We hence exploit the square wave basis, the natural output of digital systems, to design very efficient signal generators based on multiplier-less mathematics.

## *Acknowledgements*

The author wishes to thank some people whose help, support and friendship has been essential to this work: professors F. Giordano of Università Parthenope Naples, S. Cavaliere and I. Ortosecco of Università Federico II Naples, P. Parascandolo of INFN Naples, P. Corbo and G. Langwagen of Universidad ORT Montevideo Uruguay.

## *Appendix*

This article is the complement to a previous one [11] centered on vectorial analysis. There it was demonstrated that a couple of functions satisfying quite loose requisites (not including orthogonality) is a viable basis in $L^2$, hence generalizing the Fourier Theorem for vectorial decomposition, without the need for biorthogonal bases.

Here we complete the path, generalizing also the polar flavor of the Theorem to nonorthogonal bases. It is evident that a polar (or better, phase) decomposition is a more flexible tool than the vectorial analysis when nonorthogonal bases are involved, and the consequences are wide ranging. For one it is much more intuitive and easier to manage a single function as a basis. Here we shall discuss only the application to the square wave synthesizer, while a more in depth analysis and the discussion on other possible applications will be left for another paper.

Only the necessary requisites for the convergence will be given here, while the broader sufficient requisites are yet an open question. Our space of choice is that of the Lebesgue square integrable functions or $L^2$, that includes real world signals. As it is well known, it is an Hilbert space and any function of the space can be expressed as a Fourier series or transform depending if it is periodic or not.

We shall prove the feasibility of the frequency-phase (in orthogonal terms: polar) analysis on nonorthogonal bases for real periodic functions in $L^2$ $[-\pi, +\pi]$; while the extension to complex valued functions, different periods and transforms is straightforward. We split the proof in two parts: in the first part we shall prove that, when exploiting an iterative computing methodology, any set $[S(kx)] \in L^2$ $[-\pi, +\pi]$ with zero average (and *k* integer positive), spans the entire $L^2$ space (it forms a complete system). In the second part we shall determine the conditions that turn the S(kx) into a well behaved ("*converging*") basis.

Given any periodic function *f(x)* $\in L^2$ it can be expressed as Fourier series. We omit here an eventual average (a DC component) from the series as it is a simple constant that will not change our conclusions, so:

$$f(x) = \sum_{k=1}^{\infty} b_k \cos(kx + \vartheta_k) \tag{A1}$$

and given another nonzero periodic function with Fourier series:

$$S(x) = \sum_{p=1}^{\infty} s_p \cos(px + \phi_p) \tag{A2}$$

It is easy to verify that one can phase shift and amplitude scale the S(kx) in order to match any cosine component at the fundamental and at any other frequency k. Just like in the case of the square wave, we can shift and scale the entire square wave and its fundamental sinusoid (1) will be shifted and scaled accordingly. We shall than use this fundamental harmonic (a sinusoid) in our analysis algorithm and treat the rest of the spectrum of the square wave basis as a noise (at higher frequencies) that will be taken into account at subsequent iterations.

In other words we can always shift and scale the S(kx) such that its fundamental harmonic (the sinusoid at frequency *k*) can be made equal to any other sinusoid (A3). So we can use this S(kx) to reconstruct the Fourier component of *f(x)* at frequency k:

$$b_k \cos(kx + \vartheta_k) = [ M_k S( kx + \Theta_k ) ]_k \tag{A3}$$

The only problem here is that as the S(kx) is not a pure sinusoid, the reconstruction will add harmonics at multiples of k. To make things more clear let us suppose that we want to reconstruct a sinusoid by means of square waves. In the first cycle of iteration we shift and scale the square wave such that its fundamental is the sinusoid we are interested in. The result will be the given sinusoid but with noise components at frequency three, five, etc (see above). So in the subsequent iteration, to cancel the harmonic at frequency three, we subtract a square wave at frequency three and amplitude one third. And so on; see (2) and (3) above. In the fortunate case we need to reconstruct a square wave by means of square waves, only one iteration will be necessary as all the Fourier components will be fixed in one step.

This methodology can be translated into two different algorithms.

In the first, as the cosinusoid is an orthogonal function, we can separate the Fourier components at the different frequencies and solve the system of equations iteratively, starting from the fundamental, to find the unknown quantities $M_1$ and $\Theta_1$:

F1: $\quad b_1 \cos(x + \vartheta_1) = M_1 s_1 \cos(1(x + \Theta_1) + \phi_1)$ (A4)

We than equate the harmonic at frequency 2 of the $f(x)$ and the harmonic at frequency 2 of the previous reconstruction ( i.e. S(x) that is the first term on the right hand side) plus the fundamental harmonic of the $M_2 S(2x + \Theta_2)$ that is the second term on the right hand side:

F2: $\quad b_2 \cos(2x + \vartheta_2) = M_1 s_2 \cos(2(x + \Theta_1) + \phi_2) + M_2 s_1 \cos(1(2x + \Theta_2) + \phi_1)$ (A5)

Etc.

At the fundamental harmonic F1 (A4) we can separate the variables and find the unique solutions:

$$M_1 = \frac{b_1}{s_1}$$ (A6)

$$\Theta_1 = \vartheta_1 - \phi_1$$ (A7)

We can place these two values in the equation (A5) and find the new unknown $M_2$ and $\Theta_2$. Iterating the procedure at F3, F4, etc. we can reconstruct the exact Fourier components of the function $f(x)$ at any frequency by means of a series of the functions S(kx).

Another, possibly simpler, algorithm is to use at each cycle the error function, i.e. the $f(x)$ minus the previous reconstruction. At n = 1 the error function is $f(x)$ itself, at n=2 one can use the (A6) and (A7) to compute:

$f_{e2}(x) = f_{e1}(x) - M_1 S(x + \Theta_1)$ (A8)

Obviously, by construction, $f_{e2}(x)$ has only harmonics from frequency two up. One can then use the $f_{e2}(x)$ as a new function and iterate the same procedure using S(2x) to cancel the harmonic at frequency two from the error function and so on, to get:

$$f_{eN+1}(x) = f_{eN}(x) - M_N S(Nx + \Theta_N) = f(x) - \sum_{i=1}^{N} M_i S(ix + \Theta_i)$$ (A9)

So at each cycle of the iteration the lowest harmonic vanishes from the error function.

Since the error function $f_{eN+1}(x)$ in the (A9) has, by construction, no harmonics up to any frequency N, the end result will be a reconstruction of the $f(x)$ via a series of the S(kx) up to any chosen frequency N. What makes possible this result is the iterative algorithm that cancels one harmonic at a time from the Fourier spectrum of the function $f(x)$. What we did is essentially an iterative change of coordinates from the function S(kx) to the cosine and back to S(kx), see [11].

So far the (A9) proves that any function S(kx)∈$L^2$ spans the entire $L^2$ space when using the above iterative computing method to determine the coefficients.

Note that up to now the only requirements on S(x) is to have $s_1 \neq 0$ or equivalently a frequency not higher than that of the $f(x)$ and integral zero over a period.

In order to check whether S(kx) is a basis, we need to verify under which conditions the series of the S(kx) converges to $f(x)$, or that the limit of the (A9) tends to zero as N→∞. Of the different definitions of convergence: pointwise, uniform and norm, the latter appears the most useful. So we can express the previous statement more rigorously requiring that the norm of the error function tends to zero as N →∞ (following the definition of metrics in Hilbert spaces), or:

$$\lim_{N \to \infty} \| f_{eN}(x) \| = \lim_{N \to \infty} \| f(x) - \sum_{n=1}^{N} M_n S(nx + \Theta_n) \| = 0 \tag{A10}$$

We start by testing for convergence the series of the norm of the error functions. As the norm is always positive we can apply the d'Alembert's ratio test for convergence, whose requirement is the existence of an r:

$$\lim_{N \to \infty} \frac{\| f_{eN+1} \|}{\| f_{eN} \|} = r \tag{A11}$$

If r < 1 the series converges, if r > 1 the series diverges, and the test is inconclusive for r = 1. We use at each iteration of the algorithm the previous error function, so we can write $f_{eN}$ in terms of its Fourier components (just remember that by construction $f_{eN}$ has no harmonics at a frequency lower than N):

$$f_{eN} = \sum_{m=N}^{\infty} a_m \cos(mx + \delta_m) = a_N \cos(Nx + \delta_N) + \sum_{m=N+1}^{\infty} a_m \cos(mx + \delta_m) \tag{A12}$$

And the $f_{eN+1}$ will be this same function minus the component at frequency N plus some noise at multiples of N due to the term $M_N S(Nx + \Theta_N)$, see for reference (A9):

$$f_{eN+1} = a_N \cos(Nx + \delta_N) + \sum_{m=N+1}^{\infty} a_m \cos(mx + \delta_N) - M_N s_1 \cos(Nx + \Theta_N + \varphi_1) - M_N \sum_{p=2}^{\infty} s_p \cos(p(Nx + \Theta_N) + \varphi_p) \tag{A13}$$

The first and third component on the right hand side of (A13) are equal by construction, so their sum is zero (no component at frequency N). Applying the triangle inequality we get:

$$\| f_{eN+1} \| \leq \left\| \sum_{m=N+1}^{\infty} a_m \cos(mx + \delta_N) \right\| + \left\| M_N \sum_{p=2}^{\infty} s_p \cos(p(Nx + \Theta_N) + \varphi_p) \right\| \tag{A14}$$

For the same reason we can substitute the first component of (A12) with the third term of the (A13) to get:

$$\frac{\| f_{eN+1} \|}{\| f_{eN} \|} \leq \frac{\left\| \sum_{m=N+1}^{\infty} a_m \cos(mx + \delta_N) \right\| + \left\| M_N \sum_{p=2}^{\infty} s_p \cos(p(Nx + \Theta_N) + \varphi_p) \right\|}{\| M_N s_1 \cos(Nx + \Theta_N + \varphi_1) \| + \left\| \sum_{m=N+1}^{\infty} a_m \cos(mx + \delta_N) \right\|} \tag{A15}$$

And finally, the (A15) is less than 1 if the expression in the (A16) is less than 1:

$$\lim_{N \to \infty} \frac{\| f_{eN+1} \|}{\| f_{eN} \|} \leq \frac{\sum_{p=2}^{\infty} (s_p)^2}{s_1^2} = r \tag{A16}$$

So if r < 1 the series of norm error converges and consequently the norm of the error function $\|f_{eN}\|$ itself tends to zero as N →∞. Hence if the S(x) is such that the ratio r<1, the reconstruction in terms of the S(kx) tends to f(x) in norm. In the case r = 0 the basis is orthogonal and the whole theory reduces to the standard Fourier's own, as it should reasonably be expected. Thus if (A16) <1 we can finally write that, in $L^2$ norm:

$$f(x) = \sum_{n=1}^{\infty} M_n S(nx + \Theta_n) \tag{A17}$$

To prove that [S(kx)] is a basis in $L^2$ we need to prove also that the reconstruction is unique, but this comes as a direct consequence of the use of the Fourier coefficients and the uniqueness of the solution of the equations. A much faster analysis algorithm has been developed that works directly in Fourier frequency domain, having a computational complexity roughly the double of the FFT. Details will be given in a next paper.

Note that the proof is based exclusively on Fourier components so any function that can be expressed as a Fourier series can also be expressed as series of the new basis S(kx). From the (A16) results that a function is a basis if most of the energy is at the fundamental harmonic. And the square wave is precisely a member of this class of functions.

The iterative approach employed here can be reminiscent of another technique: the Empirical Mode Decomposition (EMD) [14] that also uses an iterative computation. However there are substantial differences between the two tools. For one the result of EMD is a set of orthogonal functions at different frequencies, whose sum is equal to the signal. As a matter of fact the EMD is, by definition, empirical, as one has no control on the generated bases. Instead with this new tool the basis is defined beforehand, and the result of the analysis is a set of amplitudes and phases expressing the signal in the frequency domain of the chosen basis.